\documentclass[aps,prb,twocolumn]{revtex4}
\usepackage{graphics}

\begin{document}
\title{Electric fields with ultrasoft pseudo-potentials: applications to 
benzene and anthracene}
\author{ Jaroslav T\'{o}bik  and Andrea Dal Corso}
\address{SISSA, via Beirut 2/4, 34014 Trieste, Italy.}
\address{INFM-DEMOCRITOS, National Simulation Center, Trieste, Italy.}
\date{\today}

\begin{abstract}
We present density functional perturbation theory for electric field 
perturbations and ultra-soft pseudopotentials.
Applications to benzene and anthracene molecules and surfaces are reported
as examples. We point out several issues concerning the evaluation of 
the polarizability of molecules and slabs with periodic boundary conditions. 
\end{abstract}

\maketitle

\section{Introduction}

Ultrasoft pseudo-potentials (US-PPs)~\cite{Vanderbilt} are employed
in large scale electronic structure calculations because they allow
precise and efficient simulations of localized $3d$ and $2p$ electrons 
with plane-waves basis sets.
Several ab-initio techniques have been adapted to US-PPs. Important
examples comprise the Car-Parrinello molecular dynamics,~\cite{Laasonen} 
the Berry phase approach to the macroscopic polarization 
of solids,~\cite{king-smith} 
the ballistic conductance of open quantum systems,~\cite{smogunov} 
and also density functional perturbation theory 
(DFPT).~\cite{uno,due,giapponesi} 
DFPT~\cite{baroni,rmp} addresses the response of an inhomogeneous 
electron gas to external perturbations, giving access to the lattice 
dynamics, to the dielectric, and to the elastic properties of 
materials.~\cite{baroni2}
DFPT for lattice dynamics with US-PPs has been presented in 
detail in Refs.~\onlinecite{uno,due}, whereas
the treatment of an electric field perturbation and of the 
dielectric response has been only briefly sketched 
in the literature.~\cite{giapponesi} 
The purpose of this paper is to describe our implementation of DFPT 
for electric field perturbations, to validate it with US-PPs and
to apply it to some examples. We calculate the polarizabilities 
of two molecules, benzene and anthracene, and the dielectric 
properties of the $(010)$ surface of benzene and of the $(100)$ surface 
of anthracene. 
We validate our DFPT implementation with US-PPs in two ways.
First we focus on the electronic density induced, at linear order, 
by an electric field. This density is calculated by DFPT 
and by a self-consistent finite electric field (FEF) approach. 
Second, we compare the polarizability inferred from the dipole moment 
of the induced charge and from the dielectric constant of the periodic solid 
simulated within the super-cell approach. In both cases we find a very
good agreement between DFPT and FEF.

Using plane-waves and pseudo-potentials, as implemented in present codes,
it is not possible to study truly isolated molecules or surfaces. 
A super-cell geometry must be adopted by creating 
a fictitious lattice made of periodically repeated molecules or 
slabs (with two surfaces) separated by enough vacuum and
the convergence of the calculated properties must be checked against 
the increase of the size of the vacuum region. In this respect, to 
converge the dielectric properties is particularly difficult 
because, due to the long range of electrostatic interactions, 
periodic boundary conditions may generate a spurious electric field 
which become negligible only at very large vacuum spacing. 
In this paper we show that, accounting for this spurious electric field,
it is possible to reduce considerably the vacuum needed to converge 
the polarizability both by DFPT and by FEF. 

We note in passing that in solid benzene and anthracene there are 
two distinct chemical bonds: strong covalent bonds 
within a single molecule and weak bonds between molecules responsible 
for the cohesion of the solid. Within density functional theory, 
in the local density approximation (LDA) or in the generalized gradient 
approximation (GGA), one cannot account for the van der Waals forces
which are important for molecule-molecule interactions and
therefore the geometry of the system cannot be determined from 
energy minimization.~\cite{Sprik2,Hummer}
However, once the geometry has been adjusted, the other calculated
properties often turns out to be reasonably described by LDA or GGA.
In this paper, while we allow the geometry of the isolated molecules 
to be fully relaxed in order to minimize energy, we borrow
from experiment~\cite{wyckoff} the orientations of the molecules and
the cell sizes for bulk benzene and anthracene. The geometry of the surfaces 
is assumed to be identical to the truncated bulk.

The outline of the paper is the following: Section II contains 
an expression for the dielectric constant with US-PPs.
Section III describes our implementation of the FEF approach
and other technical details. Section IV is devoted to the study 
of the polarizability of the benzene and anthracene molecules, while
in Section V the dielectric properties of slabs of benzene and anthracene 
are discussed.
 
\section{Dielectric constant with ultrasoft pseudo-potentials}
The macroscopic dielectric tensor of an insulating solid is defined 
as:~\cite{rmp} 
\begin{equation}
\epsilon_{\alpha\beta}=\delta_{\alpha\beta} + 4 \pi {d P_\alpha \over 
d E_\beta}, \label{eq1}
\end{equation}
$\alpha$ and $\beta$ are Cartesian coordinates and 
${d P_\alpha \over d E_\beta}$ is the derivative of the electronic
polarization with respect to the macroscopic screened electric 
field ${\bf E}$. 
We neglect atomic relaxations and therefore all our considerations 
regard the so called ``clamped-ions'' dielectric constant $\epsilon_{\infty}$
where only the dielectric contribution of electrons is accounted for.

In order to calculate ${d P_\alpha \over d E_\beta}$, we begin with the 
derivative of the electronic density with respect to an electric field. With
US-PPs, the electronic density is written as:
\begin{equation}
\rho({\bf r}) = \sum_i \langle \psi_{i} | K({\bf r}) | \psi_{i}\rangle
\label{rho}
\end{equation}
where the sum runs over the occupied states and the kernel
$K({\bf r}; {\bf r}_1, {\bf r}_2)$ is:
\begin{eqnarray}
K({\bf r};{\bf r}_1,{\bf r}_2)= \delta({\bf r}-{\bf r}_1)
\delta({\bf r}-{\bf r}_2)+ \nonumber \\ 
+ \sum_{Inm} Q^{\gamma(I)}_{nm}({\bf r}-{\bf R}_I)
\beta^{\gamma(I)}_n({\bf r}_1-{\bf R}_I)
\beta^{*\gamma(I)}_m({\bf r}_2-{\bf R}_I).
\label{defk}
\end{eqnarray}
The augmentation functions $Q^{\gamma(I)}_{nm}({\bf r})$ and
the projector functions $\beta^{\gamma(I)}_n({\bf r})$
are calculated togheter with the PP and are localized
about the atoms at position ${\bf R}_I$.~\cite{Vanderbilt}

The electronic charge linearly induced by an electric field is therefore:
\begin{equation}
{d n({\bf r}) \over d E_\beta} = 2 \sum_{i} \langle {d \psi_{i}
\over d E_\beta} | K({\bf r}) | \psi_{i}\rangle + c.c.,
\label{eq3}
\end{equation}
$c.c.$ means complex conjugate. 

Using Eq.~\ref{eq3}, we get ${d P_\alpha \over d E_\beta}$
as:
\begin{equation}
{d P_\alpha \over d E_\beta} = 
-{2 e\over N \Omega} \sum_{i} \int d^3 r\ 
\langle {d \psi_{i}
\over d E_\beta} | {r}_\alpha K({\bf r}) 
| \psi_{i}\rangle + c.c., 
\label{dpde}
\end{equation}
where the integral is over the volume of the solid, made up of $N$
unit cells of volume $\Omega$. Born-von K\'arm\'an boundary conditions
are assumed for the wave-functions. The electron charge is $(-e)$. 

For convenience, we define the functions
$|\phi_{i}^\alpha\rangle =  \int d^3 r\ e{r}_\alpha K({\bf r}) 
| \psi_{i}\rangle$, and introduce the projector into the
empty states manifold $P_c=\sum_c |\psi_c\rangle\langle\psi_c|S$, where
$S$ is the overlap matrix (see below). 
With these definitions, Eq.~\ref{dpde} becomes:
 \begin{eqnarray}
{d P_\alpha \over d E_\beta}= -{2 \over N \Omega} \sum_{i} 
\langle {d \psi_{i} \over d E_\beta} |P_{c}^\dagger|\phi_{i}^\alpha\rangle
+c.c..
\label{eq7}
\end{eqnarray}
The functions $|\phi_{i}^\alpha\rangle$ and hence
${d P_\alpha \over d E_\beta}$ are well defined in a finite 
system. In a periodic solid $P_{c}^\dagger|\phi_{i}^\alpha\rangle$ 
can be defined as done with norm conserving pseudo-potentials. We exploit 
the relation between the off-diagonal matrix elements of the
${\bf r}$ operator between non-degenerate Bloch states,
and the matrix elements of the velocity operator.
With US-PPs, we have:  
\begin{equation}
\langle \psi_{j} | S {r}_\alpha | \psi_{i} \rangle
= {\langle \psi_{j} | \left[ H-\varepsilon_{i}S, 
{r}_\alpha \right]  | \psi_{i} \rangle \over 
\varepsilon_{j} -\varepsilon_{i} },
\label{eq8}
\end{equation}
where $H$ is the unperturbed Hamiltonian and $\varepsilon_{i}$ are 
the unperturbed eigenvalues.
The overlap matrix $S$ is~\cite{Vanderbilt} 
\begin{widetext}
\begin{equation}
S({\bf r}_1,{\bf r}_2)=\delta({\bf r}_1-{\bf r}_2)
+ \sum_{Inm} q^{\gamma(I)}_{nm}
\beta^{\gamma(I)}_n({\bf r}_1-{\bf R}_I)
\beta^{*\gamma(I)}_m({\bf r}_2-{\bf R}_I), 
\label{defs} 
\end{equation}
where the coefficients $q^{\gamma(I)}_{nm}=\int {\it d^3r}\ 
Q^{\gamma(I)}_{nm}({\bf r})$ are the integrals of the augmentation
functions. 

Using Eq.~\ref{eq8}, we see that $P_{c} {r}_\alpha 
| \psi_{i} \rangle$ are the solutions of the linear system:
\begin{equation}
\left[ H+Q - \varepsilon_{i}S
\right] P_{c} {r}_\alpha | \psi_{i} \rangle = P_{c}^\dagger \left[ H-\varepsilon_{i}S, 
{r}_\alpha \right] | \psi_{i} \rangle,
\label{eq9}
\end{equation}
where both the left and the right hand sides are lattice periodic. 
$Q$ is added to the linear system in order to make it non singular
as explained in detail in Ref.~\onlinecite{rmp} (See Eq.~30 and Eq.~72). 
With norm conserving pseudopotentials, in insulators, $Q$ is proportional 
to the valence band projector. Its generalization to US-PPs is given
in Ref.~\onlinecite{due} (see discussion after Eq.~29).
By solving this linear system, we obtain 
$P_{c} {r}_\alpha | \psi_{i} \rangle$, while we need
$ P_{c}^\dagger {r}_\alpha | \psi_{i} \rangle$
to compute $P_{c}^\dagger|\phi_{i}^\alpha\rangle$.
Since $S P_{c} =P_{c}^\dagger S$, we have $S P_{c} 
{r}_\alpha | \psi_{i} \rangle  = P_{c}^\dagger S 
{r}_\alpha | \psi_{i} \rangle$, hence the functions $P_{c}^\dagger
 |\phi^\alpha_i\rangle$
are:
\begin{equation}
P_c^\dagger |\phi_i^\alpha \rangle = S P_c e{r}_\alpha | \psi_i \rangle - 
P_c^\dagger \sum_{Inm} q^{\gamma(I)}_{nm} |\beta_n^I \rangle 
\langle \beta_m^I | e{r}_\alpha| \psi_i \rangle +
P_c^\dagger \sum_{Inm}{ 
I^\alpha_{Inm}}
|\beta_n^I \rangle \langle \beta_m^I |
\psi_i \rangle,
\label{defphi}
\end{equation}
where we defined the integral $I^\alpha_{Inm}= \int d^3 r\ e{r}_\alpha
Q^{\gamma(I)}_{nm}({\bf r}-{\bf R}_I)$.

To proceed further, we need the first order derivative of the electronic 
wave-functions $|\psi_{i}\rangle$ with respect to an electric field. We 
can calculate these quantities to linear order in perturbation theory.
The overlap matrix $S$ does not depend on the electric field while the 
differentiation of the Kohn and Sham potential yields:
\begin{equation}
{d V_{KS} \over d E_\beta} =\int d^3 r \left[ e{r}_\beta + 
{d V_{Hxc}({\bf r}) \over d E_\beta} \right] K({\bf r}),
\label{eq5}
\end{equation}
where $V_{Hxc}$ is the Hartree and exchange and correlation potential.
Applying $P_{c}^\dagger$ to Eq.~(19) of Ref.~\onlinecite{due}, we obtain:
\begin{equation}
\left[ H + Q - \varepsilon_i S \right] 
P_{c}|{d \psi_{i}\over d E_\beta} \rangle 
= -P_{c}^\dagger \int d^3 r \left[ e{r}_\beta + {d V_{Hxc}({\bf r}) 
\over d E_\beta} \right] K({\bf r}) |\psi_{i} \rangle.
\label{eq6}
\end{equation}
\end{widetext}
The self-consistent solutions of this linear system, together with 
Eq.~\ref{defphi}, are substituted into Eq.~\ref{eq7} to calculate 
${d P_\alpha \over d E_\beta}$.
Finally, the dielectric tensor is calculated via Eq.~\ref{eq1}.
\section{Technical details}

We validate the theory by comparison with a self-consistent FEF method. 
A FEF is simulated as suggested by Kunc and Resta:~\cite{KR1,KR2} 
a sawtooth-like potential is added to the bare ionic potential.
This method, when applied to periodic systems, is not as 
efficient as other recent 
approaches~\cite{Nunes,Fernandez,Vanderbiltef,Pasquarello,Umari} 
because it requires the simulation of a super-cell, but its implementation 
is simple. Our systems, molecules and slabs, require already
a super-cell and the method of Kunc and Resta suits adequately our goals.
The molecules and the slabs are inserted in the region where 
the sawtooth-like potential is linear like the 
potential of an electric field (${\bf E}_{sl}$):
$\Phi({\bf r})=-{\bf E}_{sl}\cdot {\bf r}$. To ensure
periodicity, the slope of the potential is reversed in a small region 
in the middle of the vacuum. 
For finite super-cells and small enough fields, the electrons are 
only slightly polarized by the field, the systems maintain a 
well defined energy gap between occupied and empty electronic states
and electrons do not escape into the vacuum. 
In these simulations, the macroscopic electric field ${\bf E}$ is zero 
because, solving the Poisson equation we assume, as a boundary
condition, the periodicity of the Hartree potential.
The sawtooth-like potential describes a microscopic electric field ${\bf E}_{sl}$ 
which vanishes averaging over macroscopic distances. 
On our finite systems, this microscopic field has the same 
effect of a macroscopic electric field. Indeed, in DFPT, as derived in 
previous Section, the perturbation is a genuine macroscopic electric 
field ${\bf E}$ and, as shown in the following, the electronic density 
induced, at linear order, by the field ${\bf E}_{sl}$ and the electronic 
density induced by the macroscopic electric field ${\bf E}$ 
are equal to each other when ${\bf E}_{sl}={\bf E}$.

Both the FEF approach and DFPT for US-PPs have been implemented 
in the {\tt PWscf}~\cite{PWSCF} package. 
All calculations are carried out within the GGA approximation.
The expression of the exchange and correlation energy introduced by 
Perdew, Burke and Ernzerhof (PBE)~\cite{PBE} is used in the GGA
functional. The US-PPs of hydrogen and carbon have the 
parameters described in Ref.~\onlinecite{due}. 
Plane waves up to an energy cutoff of 30 Ry for the wave-functions
and 180 Ry for the electron density, are used.
Only the $\Gamma$ point is used for the Brillouin zone (BZ) sampling 
in the molecular calculation while a $3\times 3$ Monkhorst and Pack 
mesh~\cite{MP} of ${\bf k}$-points is used for sampling the two-dimensional 
BZ of the slabs.

\section{Molecules}
\begin{figure}[t]
{\par\centering \resizebox*{0.5\textwidth}{!}{\rotatebox{0}{\includegraphics{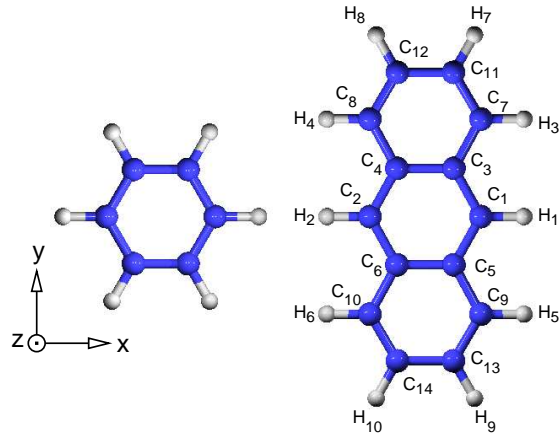}}}\par}
\caption{Benzene and anthracene molecules. Distances and angles are 
reported in Tables \ref{tab-bgeom} and \ref{tab-ageom} for benzene
and anthracene respectively.  The polarizability tensors are given with
respect to the axes shown in the figure.}
\label{fig-mol}
\end{figure}
\begin{figure*}[t]
{\par\centering \resizebox*{1.0\textwidth}{!}{\rotatebox{270}{\includegraphics{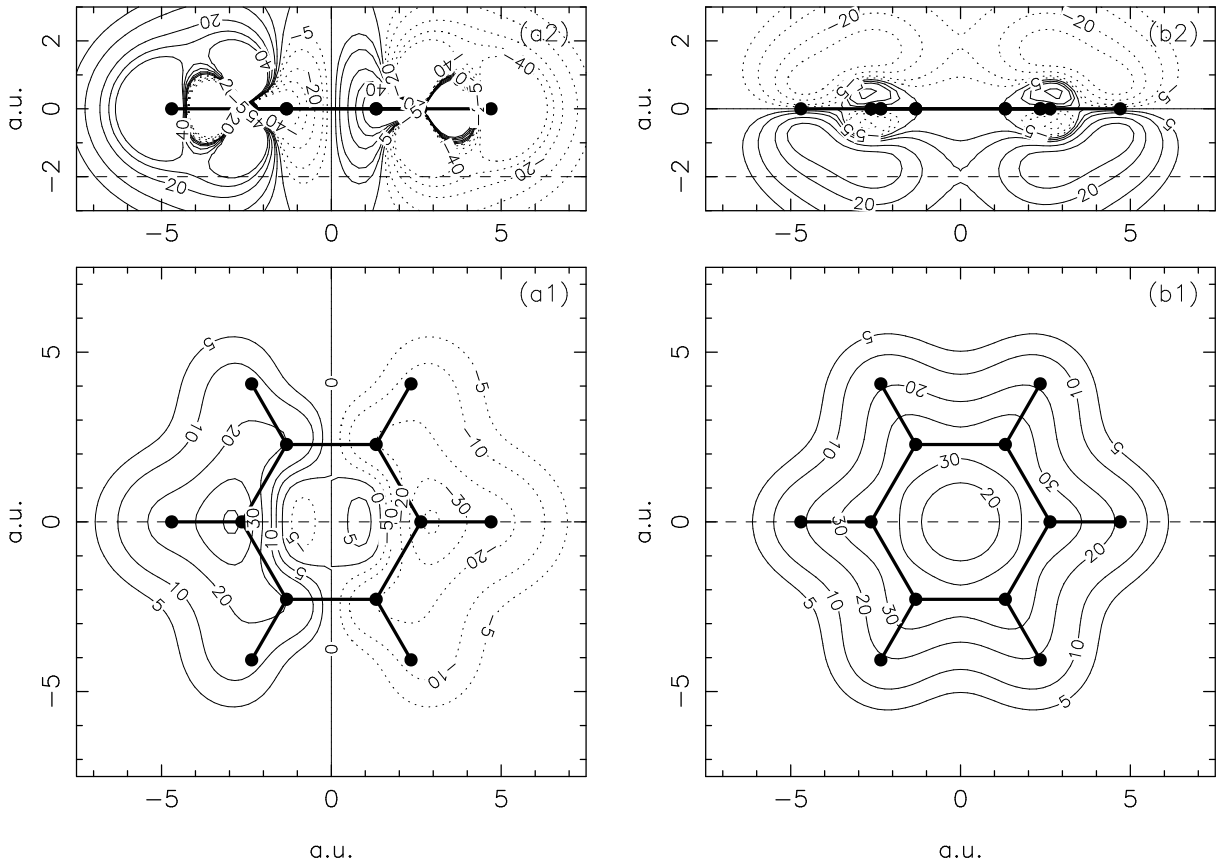}}}\par}
\caption{Electronic density induced by an electric field either 
parallel (a) or perpendicular (b) to the plane of benzene calculated by DFPT. 
(1) Plane perpendicular to the $z$ axis passing through 
$(0.0,0.0,-2.0)$ a.u.; (2) plane $xz$.
Contours are in units of $10^{-3}$ elec./(a.u.)$^3$.
The contours correspond to a field $0.5$ a.u. ($2.57\times 10^{9}$ V/cm).}
\label{fig-bench}
\end{figure*}
\begin{table}[b]
\begin{tabular}{cccc}
     &\begin{tabular}{c} this work\\ (PBE) \end{tabular}&\begin{tabular}{c} Ref.~\onlinecite{JCP-Dele} \\(B3LYP) \end{tabular}& 
\begin{tabular}{c}Ref.~\onlinecite{wyckoff}\\ (expt.) \end{tabular}\\ 
\hline
$C-C$ & 1.396 & 1.399 & 1.398 (n), 1.392 (x)\\ 
$C-H$ & 1.091 & 1.092 & 1.090 (n)  \\ 
\end{tabular}
\caption{Theoretical GGA optimized geometry of the benzene molecule
compared with previous theoretical work~\cite{JCP-Dele} (B3LYP with 
localized basis set) and experiment.~\cite{wyckoff} Bond lengths 
are in \AA . The symbols are defined in Fig.~\ref{fig-mol}. Abbreviations: 
(n) neutron, (x) x-ray diffraction.}
\label{tab-bgeom}
\end{table}
\begin{figure*}[t]
{\par\centering \resizebox*{1.0\textwidth}{!}{\rotatebox{270}{\includegraphics{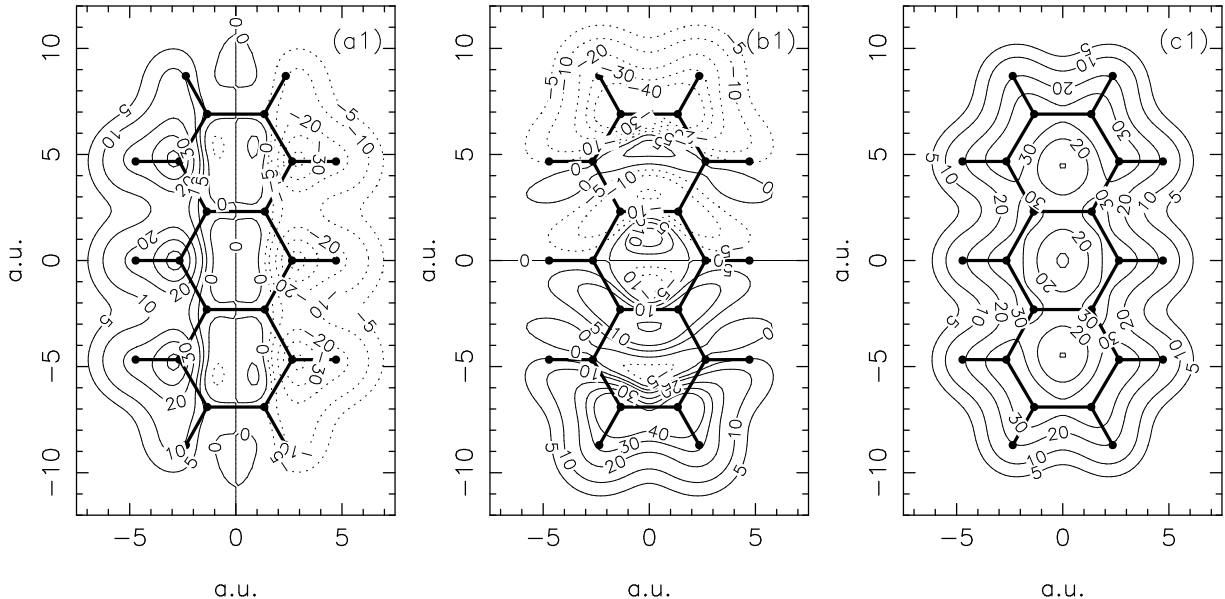}}}\par}
\caption{Electronic density induced by an electric field either 
parallel (a,b) or perpendicular (c) to the plane of anthracene calculated 
by DFPT. 
In (a) the field is parallel to the short axis, in (b) it is parallel
to the long axis of the molecule.
The plane is perpendicular to the $z$ axis passing through 
the point $(0.0,0.0,-2.0)$ a.u.. 
Contours are in units of $10^{-3}$ elec./(a.u.)$^3$.
The contours correspond to a field $0.5$ a.u..} 
\label{fig-antch}
\end{figure*}
\begin{table}[b]
\begin{tabular}{ccccc}
bond & \begin{tabular}{c}this work\\ (PBE)\end{tabular} & \begin{tabular}{c}Ref.~\onlinecite{JCP-Dele}\\ (B3LYP) \end{tabular}&\begin{tabular}{c} Ref.~\onlinecite{Ketkar}\\ (expt.)\end{tabular} & \begin{tabular}{c}Ref.~\onlinecite{wyckoff}\\ (expt. solid)\end{tabular} \\
\hline 
$C_1-C_3$       &    1.398 & 1.403 & 1.392 & 1.403 \\
$C_3-C_7$       &    1.425 & 1.432 & 1.437 & 1.445 \\
$C_7-C_{11}$    &    1.370 & 1.372 & 1.397 & 1.374 \\
$C_3-C_4$       &    1.446 & 1.447 & 1.437 & 1.425 \\
$C_{11}-C_{12}$ &    1.421 & 1.428 & 1.422 & 1.412 \\
$C_1-H_1$       &    1.092 & 1.094 & -     & 1.121 \\
$C_{11}-H_{7}$  &    1.090 & 1.092 & -     & 1.139 \\
$C_7-H_3$       &    1.091 & 1.093 & -     & 1.153 \\
\hline
angle     &          &       &       \\
\hline
$C_3C_1C_5       $ &  121.74 & 121.80 & - & 120.37 \\
$C_3C_7C_{11}    $ &  120.89 & 120.97 & - & 120.31 \\
$H_3C_7C_{11}    $ &  120.55 & 120.53 & - & 123.48 \\
$H_7C_{11}C_{12} $ &  119.34 & 119.43 & - & 124.93 \\
\end{tabular}
\caption{Theoretical GGA optimized geometry of the anthracene molecule 
compared with previous theoretical work~\cite{JCP-Dele} 
(B3LYP with localized basis set) and
experiment.~\cite{Ketkar} The experimental geometry of solid anthracene
is also reported for comparison.~\cite{wyckoff} 
Bond lengths are in \AA , angles in degrees. The symbols are defined 
in Fig.~\ref{fig-mol}.}
\label{tab-ageom}
\end{table}
\begin{figure}[t]
{\par\centering \resizebox*{0.5\textwidth}{!}{\rotatebox{0}{\includegraphics{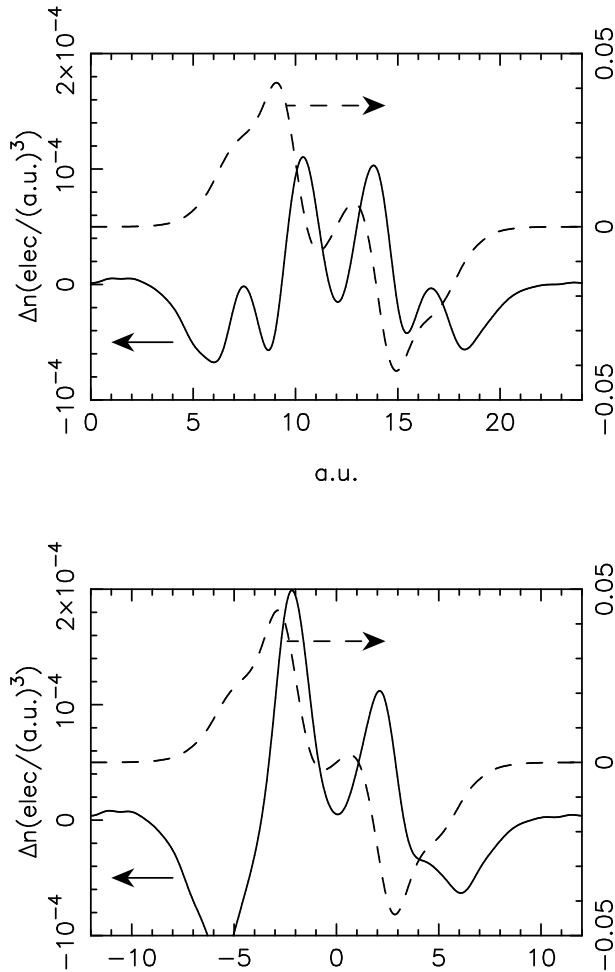}}}\par}
\caption{Density profiles along the line 
($\lambda$,0.0,$-$2.0) a.u., $-$12 a.u.$<\lambda<$12 a.u., (dashed line) 
and the difference between DFPT and FEF (continuous line). Upper panel 
for benzene in a cubic super-cell with size 24 a.u.
The lower panel for anthracene in an orthorhombic super-cell
with sizes 24 a.u. $\times$ 36 a.u. $\times$ 15 a.u..}
\label{fig-diff}
\end{figure}
\begin{table*}[t]
\begin{tabular}{p{0.10\textwidth}p{0.12\textwidth}p{0.12\textwidth}p{0.12\textwidth}p{0.12\textwidth}p{0.12\textwidth}}
 & Benzene &  & & Anthracene &   \\
$L$ (a.u.)& $\alpha_{xx}$ & $\alpha_{zz}$ & $\alpha_{xx}$ & $\alpha_{yy}$ 
& $\alpha_{zz}$ \\
\hline 
 20 & 83.6 (87.4) & 44.7 (45.7) & --- &  --- & --- \\
 24 & 83.5 (85.7) & 44.9 (45.5) & 170 (179) & 317 (351) &  85.5 (87.8) \\
 28 & 83.5 (84.8) & 44.9 (45.3) & 171 (177) & 313 (333) &  86.3 (87.8) \\
 30 & 83.5 (84.6) & 45.0 (45.3) & 172 (176) & 310 (327) &  86.6 (87.9) \\
 32 & 83.4 (84.3) & 45.0 (45.2) & 172 (176) & 309 (322) &  86.8 (87.8) \\
 50 & ---         & ---         & 172 (173) & 306 (309) &  86.5 (86.8) \\
\end{tabular}
\caption{Independent components of the molecular polarizability tensor of 
benzene and anthracene (in a.u.) calculated from the dipole moment of 
$-e\Delta n$ including Lorentz field corrections. Values in parenthesis 
are calculated neglecting Lorentz field corrections.}
\label{tab-polar}
\end{table*}
\begin{table*}[!]
\begin{tabular}{p{0.10\textwidth}p{0.12\textwidth}p{0.12\textwidth}p{0.12\textwidth}p{0.12\textwidth}p{0.12\textwidth}}
 & Benzene &  & & Anthracene &   \\
$L$ (a.u.)& $\epsilon_{xx}\ (\alpha_{xx})$ & $\epsilon_{zz}\ (\alpha_{zz})$
             & $\epsilon_{xx}\ (\alpha_{xx})$ & $\epsilon_{yy}\ (\alpha_{yy})$ 
             & $\epsilon_{zz}\ (\alpha_{zz})$ \\
\hline \\
 20 &1.1372 (83.5) &1.0716 (44.5) & --- & --- & --- \\
 24 &1.0778 (83.4) &1.0412 (44.7) &1.1625 (170)&1.3329 (330)& 1.0793 (85.0) \\
 28 &1.0485 (83.3) &1.0259 (44.8) &1.1013 (171)&1.1908 (313)& 1.0501 (86.1) \\
 30 &1.0393 (83.3) &1.0210 (44.8) &1.0819 (171)&1.1519 (311)& 1.0407 (86.3) \\
 32 &1.0323 (83.3) &1.0173 (44.9) &1.0673 (171)&1.1233 (309)& 1.0335 (86.5) \\
\end{tabular}
\caption{Dielectric constant of benzene and anthracene molecules in a cubic
super-cell as a function of the box size. In parenthesis we report 
the polarizability (in a.u.) obtained from Eq.~\ref{eq-cm}.}
\label{tab-diel}
\end{table*}
\begin{table}[t]
\begin{tabular}{c c c c l}
$\alpha_{xx}$ & $\alpha_{yy}$ & $\alpha_{zz}$ & $\bar \alpha$ & Method/reference  \\
\hline
\multicolumn{3}{l}{Anthracene} & &   \\
164 & 287 & 81.2 & 177 & APSC + B3LYP/6-311G(d,p) \cite{Soos} \\
158 & 266 & 81.8 & 167 & APSC + MP2 corrections \cite{Soos} \\
172 & 303 & 86.5 & 187 & APSC + TZVP-FIP \cite{Soos} \\
171 & 306 & 86.8 & 188 & this work (GGA-PBE) \\
191 & 337 & 96.2 & 208 & DFT with TZVP-FIP basis \cite{Reis} \\
169 & 240 & 105  & 171 & EFNMR - BS (static)\cite{pol-nmr} \\
173 & 238 & 103  & 171 & KE, EP, CME - BS (dynamic)\cite{Cheng-AJC} \\
165 & 242 & 107  & 171 & KE, EP, EST - BS (dynamic)\cite{LeFevre} \\
174 & 272 & 80.3 & 175 & CR (dynamic)~\cite{Vuks} \\
139 & 292 & 115  & 182 & CR (dynamic)~\cite{Bounds-ChP} \\
\multicolumn{3}{l}{Benzene}     & &   \\
74.2  & & 39.5 & 62.6 & HF \cite{pol-mp2} \\
76.0  & & 41.3 & 64.4 & HF + MP2 \cite{pol-mp2} \\
76.8  & & 37.2 & 63.6 & HF with DZP' basis \cite{Perez-JMS} \\
78.2  & & 37.2 & 64.5 & as above + MP2 \cite{Perez-JMS} \\
79.5  & & 45.2 & 68.1 & HF with POL basis \cite{Perez-JMS} \\
81.6  & & 45.2 & 69.5 & as above + MP2 \cite{Perez-JMS} \\
83.4  & & 45.0 & 70.6 & this work (GGA-PBE) \\
83.7  & & 44.9 & 70.9 & LDA, Gaussian basis \cite{pol-dft} \\
85.0  & & 45.6 & 72.2 & as above - different basis \cite{pol-dft} \\
74.9  & & 49.9 & 66.6 & Ref. \onlinecite{LeFevre} \\
74.9  & & 50.6 & 66.8 & KE, EP, CME (dynamic) \onlinecite{Cheng-AJC} \\
\end{tabular}
\caption{Theoretical and experimental values of benzene and 
anthracene polarizabilities. All values are in atomic units. Abbreviations: 
Atomic polarizability in a self-consistent local field (APSC), benzene
solution (BS), Kerr effect (KE), Cotton-Mouton effect (CME),
electron polarization (EP), crystal refraction (CR), empirical estimate (EST),
Hartree-Fock (HF), M{\o}ller-Plesset correction of 2nd order (MP2).}
\label{tab-exp}
\end{table}

Benzene (C$_6$H$_6$) and anthracene (C$_{14}$H$_{10}$) are planar 
molecules, their geometries are shown in Fig.~\ref{fig-mol}.
We optimize the geometries and our theoretical bond lengths 
and angles are reported in the Tabs.~\ref{tab-bgeom} and ~\ref{tab-ageom}
and compared with previous calculations~\cite{JCP-Dele} and 
with experiment.~\cite{Ketkar}
Overall, the agreement between theory and experiment is good and
our values compare well also with the more precise 
B3LYP results.~\cite{JCP-Dele}

As a first test of DFPT, we calculate the electronic density 
$\Delta n={dn\over dE}E$ induced at linear order by an electric field. 
Fig.~\ref{fig-bench} (Fig.~\ref{fig-antch}) shows
the density induced in benzene (anthracene) by an electric field either
parallel or perpendicular to the molecular plane.
The same $\Delta n$ has been calculated by FEF via a numerical 
differentiation of the self-consistent density with 
$|{\bf E}_{sl}|=10^{-3}$ a.u. and $|{\bf E}_{sl}|=0$ a.u.. 
In all cases, to the scale of the figures, the $\Delta n$ calculated by 
DFPT and the $\Delta n$ calculated by FEF coincide. In 
Fig.~\ref{fig-diff} we report, as an example, the difference 
between two $\Delta n$ to an enlarged scale. 
This error is due to nonlinear effects (present in the FEF results 
but not in DFPT) as well as to numerical noise.
Its magnitude, lower than 1\% of $\Delta n$, is similar 
with norm-conserving PPs.

Now we can address the molecular polarizabilities.
The polarizability of a molecule is a tensor which measures, at
linear order, the tendency of the molecule to change its dipole moment 
when inserted into an electric field. It is defined by the
relationship ${\bf p}_\alpha = \sum_\beta \alpha_{\alpha\beta} 
{\bf E}_{loc,\beta}$,
between the dipole moment ${\bf p}$ of the induced charge density, and
${\bf E}_{loc}$ the electric field acting on the molecule.
In general, the polarizability tensor $\alpha$ can be made diagonal in
the principal axes. 
The point group of benzene, $D_{6h}$, and of anthracene, $D_{2h}$, are
centrosymmetric, hence these two molecules have no permanent dipole moment. 
Their principal axes are shown in Fig.~\ref{fig-mol}.
In these axes, $\alpha$ has two independent components in 
benzene ($\alpha_{xx}=\alpha_{yy}$ parallel 
and $\alpha_{zz}$ perpendicular to the molecular plane), and three independent 
components in anthracene ($\alpha_{xx}$, $\alpha_{yy}$, $\alpha_{zz}$, 
parallel to the short molecular axis, parallel to the long
molecular axis and perpendicular to the molecular plane, respectively). 
We first extract the polarizabilities from the dipole moment of the
induced charge density $-e \Delta n$. 
As shown in Figs.~\ref{fig-bench} and \ref{fig-antch}, 
the $\Delta n$ induced in our molecules are localized and well separated 
by their periodic images, so that ${\bf p}$ can be calculated by 
a numerical integration (${\bf p}=-e\int 
\Delta n({\bf r}){\bf r}\ d^3 r$) over the super-cell volume.
The small differences between induced charges calculated by DFPT and by FEF  
lead to differences smaller than $1 \%$ in the dipole moments 
(for instance, in benzene, for $L=24$ a.u. and $E=0.5$ a.u., $p_x=42.79$ a.u. 
with DFPT and $p_x=42.85$ a.u. with FEF). The polarizabilities differ also 
by less than $1 \%$ because, at linear order, ${\bf E}_{loc}$ 
acting on the molecules is the same.
Indeed, the local field can be estimated recognizing that the periodic
solid simulated in the super-cell approach, has a macroscopic polarization
${\bf P}={\bf p}/\Omega$. Hence, the local field 
which acts on the molecules is given by the Lorentz formula
${\bf E}_{loc}= 
{\bf E}_{sl}+{4\pi\over 3} {\bf P}$ (FEF)
or ${\bf E}_{loc}= {\bf E}+{4\pi\over 3} {\bf P}$ (DFPT).
Of course, these formulas are valid for isotropic solids 
and cubic super-cells, however the isotropic formula
is sufficient to correct the main effects of the
periodic boundary conditions when the electric field is parallel to
one principal axis and the super-cell is cubic.
We evaluate the dipole moment of benzene 
and anthracene for several sizes of the box.
In Tab.~\ref{tab-polar}, we report the polarizabilities calculated 
with or without the Lorentz corrections.
In benzene, with a super-cell of $20$ a.u. and a polarizability 
$\alpha_{xx}=83.6$ a.u., the local field is $4.4 \%$ larger 
than $E_{sl}$ or $E$. In our largest super-cell ($L=32$ a.u.) the 
local field is still $1 \%$ larger than $E_{sl}$ or $E$. 
From Tab.~\ref{tab-polar} we can see that the inclusion of
the Lorentz correction increases the convergence rate of $\alpha_{xx}$ and
$\alpha_{zz}$. A box of $20$ a.u. is sufficient to give values 
converged within $1 \%$. 
The anthracene molecule is about $18$ a.u. long in the $y$ direction 
and therefore the molecules can be considered as truly isolated
only for super-cell sizes larger than $28$ a.u.. 
At smaller box sizes, not only electrostatic, but also direct 
molecule-molecule interactions are important.
At $L=28$ a.u., including the Lorentz correction,
$\alpha_{xx}$ and $\alpha_{zz}$ are already converged within $1 \%$. 
$\alpha_{yy}$ is instead more difficult to converge.
A super-cell of about $50$ a.u. is needed to reduce the local field 
effects below $1\%$. Instead, including Lorentz corrections, a box 
size of $32$ a.u. is sufficient to converge $\alpha_{yy}$ within $1 \%$.

Besides the direct approach, molecular polarizabilities can be evaluated 
also starting from the dielectric constant (Eq.~\ref{eq1}) of 
the periodic solid simulated in the super-cell approach.
The anisotropic Clausius-Mossotti formula which derives
from the approximate Lorentz field given above is:~\cite{an-cm}
\begin{equation}
\alpha_{\alpha\alpha} = {3 \Omega \over 4\pi} 
{\epsilon_{\alpha\alpha}-1\over \epsilon_{\alpha\alpha}+2}.
\label{eq-cm}
\end{equation}
In Tab.~\ref{tab-diel}, we report the dielectric constant 
as a function of the super-cell size and the polarizabilities
calculated via Eq.~\ref{eq-cm}, for both molecules.
The convergence rate of the polarizabilities is similar to the 
convergence rate found in Tab.~\ref{tab-polar} including the Lorentz field. 
In all cases the difference of the final polarizabilities in  
Tab.~\ref{tab-polar} and in Tab.~\ref{tab-diel} is below $1 \%$.

The final calculated components of the benzene polarizabilities are 
$\alpha_{xx}=83.4$ a.u. and $\alpha_{zz}=45.0$ a.u..
These values give a mean polarizability 
$\bar \alpha=\frac{1}{3}(2\alpha_{xx}+\alpha_{zz})=70.6$ a.u. and
are in good agreement with the results of previous theoretical
works and with experiment. We report in Tab.~\ref{tab-exp} previous
theoretical data and experimental values. 
For anthracene we get $\alpha_{xx}=171$ a.u., $\alpha_{yy}=306$ a.u.,
$\alpha_{zz}=86.8$ a.u.. These values give a mean polarizability
$\bar \alpha = 188$ a.u.. Experimental values of the anisotropic 
components of the polarizability of anthracene exist for both 
diluted benzene solutions and for solid anthracene. The values 
of $\alpha_{xx}$ ranges from $139$ a.u. to $174$ a.u., 
$\alpha_{yy}$ between $238$ a.u. and $292$ a.u. 
and $\alpha_{zz}$ between $80$ a.u. and $115$ a.u. 
(see Tab.~\ref{tab-exp}). Note that the ionic contribution to the
polarizability, which is not accounted for in our calculation, is
present in the experiment of Ref.~\onlinecite{pol-nmr} but not in 
those of Refs.~\onlinecite{Cheng-AJC,LeFevre}.
However it has been estimated that the 
ionic contribution is only of the order of $5 \%$.~\cite{pol-nmr} 
\section{Surfaces}
\begin{figure}
{\par\centering \resizebox*{0.45\textwidth}{!}{\rotatebox{0}{\includegraphics{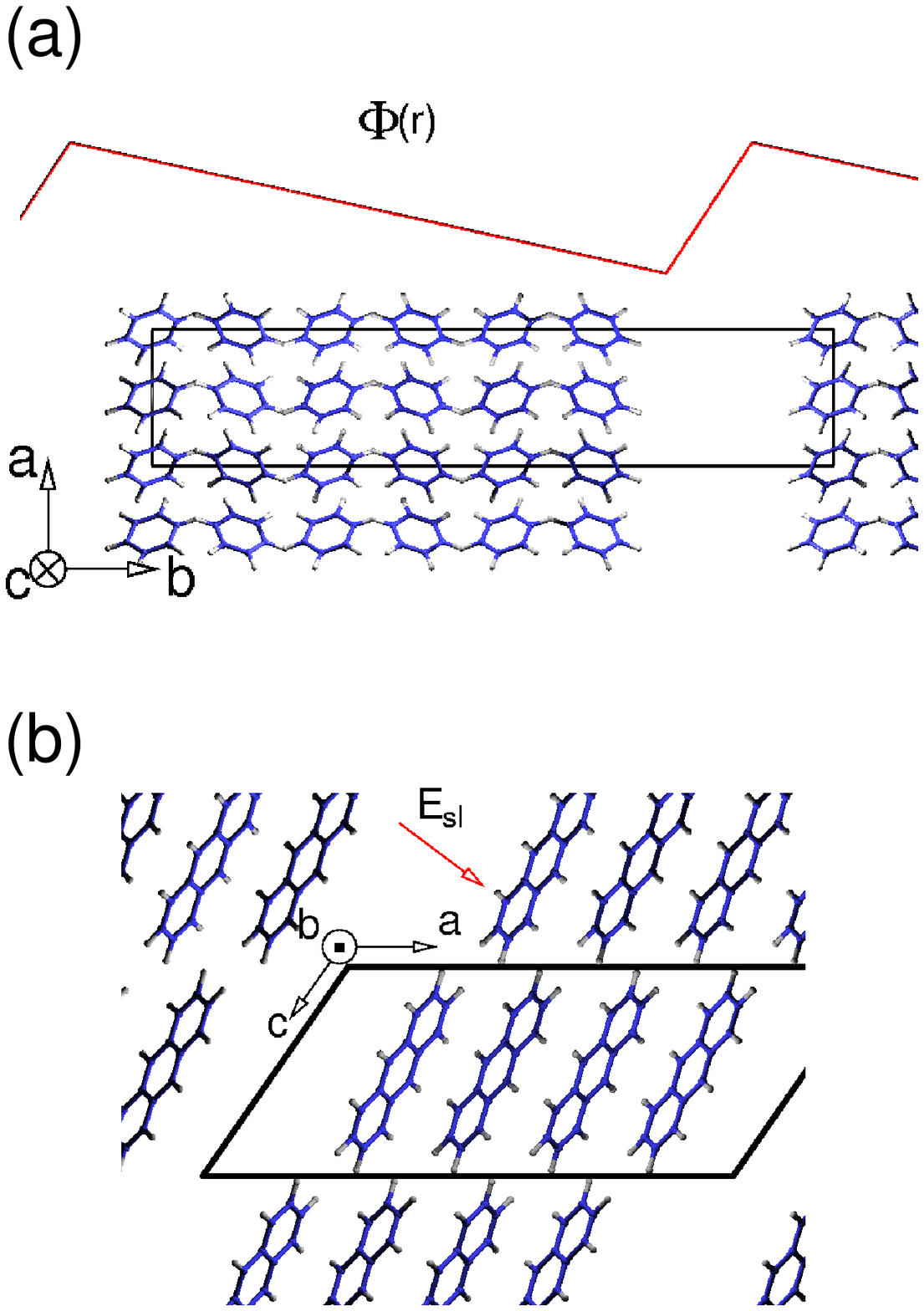}}}\par}
\caption{Geometry of the benzene (a) and anthracene (b) slabs. 
The unit cell is shown by a black frame. In (a) the applied sawtooth--like 
potential is also indicated. 
}
\label{fig-slabs}
\end{figure}
\begin{figure}[t]
{\par\centering \resizebox*{0.5\textwidth}{!}{\rotatebox{0}{\includegraphics{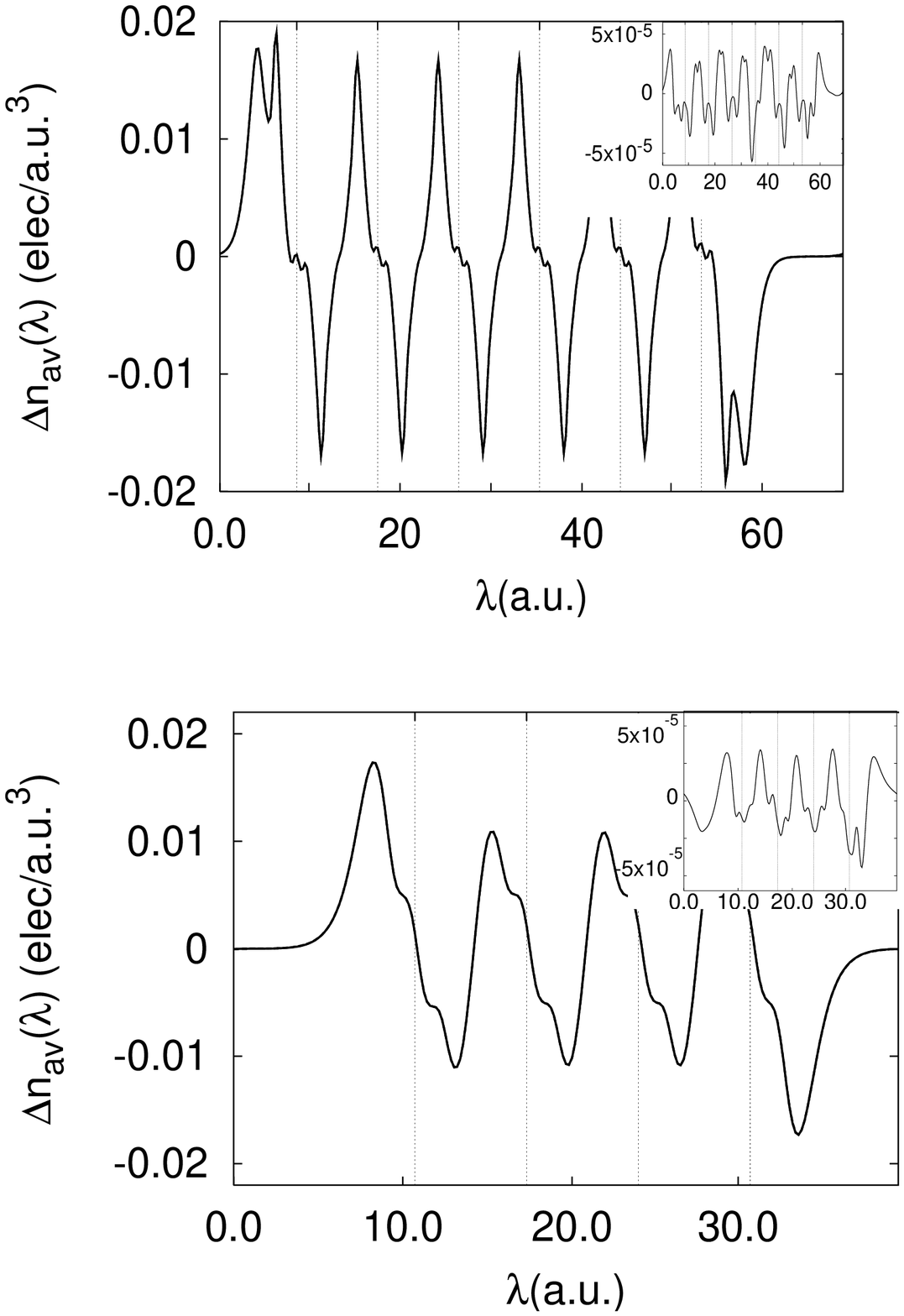}}}\par}
\caption{
Planar average $\Delta n_{av}$ of the electron density induced by an 
electric field on the benzene slab (upper panel) and anthracene slab 
(lower panel). The difference between DFPT and FEF is shown in the insets.
$\Delta n$ corresponds to an electric field of $0.5$ a.u.. Vertical lines
indicate the position of the centers of the molecules in each layer. 
}
\label{fig-chtravg}
\end{figure}

\begin{figure}[t]
{\par\centering \resizebox*{0.5\textwidth}{!}{\rotatebox{270}{\includegraphics{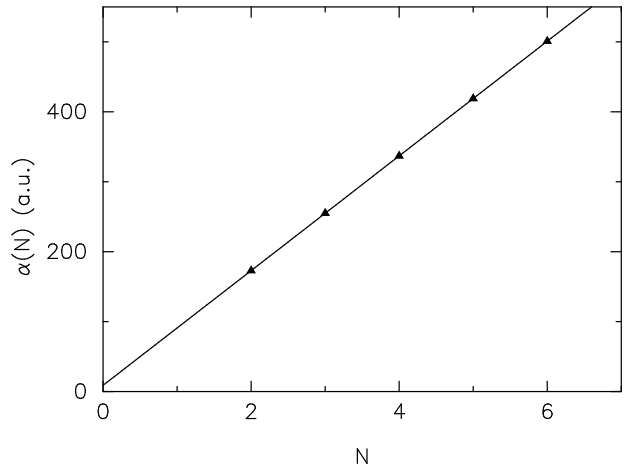}}}\par}
\caption{
Benzene slab polarizability as a function of the number of layers. The data
are interpolated with a linear fit (see Eq.~\ref{final}).
}
\label{fig-pol-lay}
\end{figure}
\begin{table}[t]
\begin{tabular}{cccccc}
 & Benzene & & & Anthracene &   \\
$L$ (a.u.)& $\alpha$ (Eq.~\ref{alpha}) & $\alpha$ 
&$L$ (a.u.)& $\alpha$ (Eq.~\ref{alpha}) & $\alpha$ \\
\hline 
65.0 &  500.7   & 1116   & 36.17 & 298.6 & 523.9 \\
68.9 &  500.9   & 1044   & 39.46 & 298.0 & 491.3 \\
75.0 &  500.9   & 959.8  & 42.75 & 298.1 & 468.2 \\
80.0 &  500.9   & 907.9  & 46.03 & 298.1 & 449.9 \\
85.0  & 500.9   & 866.5  & 49.32 & 298.1 & 435.1 \\
\end{tabular}
\caption{Polarizability of the benzene and anthracene slabs (in a.u.) 
as a function of the size of the super-cell. Values corrected for the
electric field due to the periodic boundary conditions and uncorrected
values are both reported. }
\label{tab-pol-slab}
\end{table}

Some surfaces of molecular crystals can be cleaved by cutting only weak 
molecule-molecule bonds. Such surfaces are expected to remain 
insulating and to have dielectric properties similar to the bulk. 
In this section, we present the dielectric properties of two of these
surfaces: the $(010)$ surface of benzene and the $(100)$ surface of anthracene.
Moreover, we show how to calculate the dielectric properties of 
a single slab from the dielectric constant of the solid simulated in 
the super-cell approach.

The $(010)$ benzene surface is simulated by a six-layers slab and two molecules 
per layer (see Fig.~\ref{fig-slabs}a). The super-cell is orthorhombic 
with sizes $13.78$~a.u.\ $\times L$~a.u.\ $\times 12.74$ a.u., where
$L$ depends on the vacuum between slabs. 
The $(100)$ surface of anthracene is described by a four-layers
slab and one molecule per layer (see Fig.~\ref{fig-slabs}b).
The super-cell is monoclinic with sizes 
$|{\bf a}|=L/sin\gamma$, $|{\bf b}|= 11.41$~a.u.\ and $|{\bf c}|=21.14$~a.u., 
where the angle between ${\bf a}$ and ${\bf c}$ is $\gamma=124.7^\circ$.
The long axis of the molecules is approximately parallel to the ${\bf c}$ 
vector.  
Fig.~\ref{fig-slabs}a shows the shape of the sawtooth-like potential 
used in the FEF simulations. 
The applied field is normal to the surface, in the direction
$(010)$ in benzene and in the direction of ${\bf b} \times {\bf c}$  
in anthracene. The band structures of these slabs have been reported 
in Ref.~\onlinecite{noi}.

We begin by a comparison of the induced electronic density calculated by
DFPT and by FEF. In order to visualize the induced density, we introduce
the planar average of $\Delta n$, defined as $\Delta n_{av}(\lambda)=
{1/S} \int_S \Delta n({\bf r})\ dS$. Here $\lambda$ is a coordinate along 
the surface normal and $S$ is the area of one surface unit cell. The 
integration is on cross sections parallel to the surface. 
In Fig.~\ref{fig-chtravg}, we report $\Delta n_{av}$ calculated 
by DFPT and by FEF.
In the latter case the numerical differentiation is done taking the 
field $E_{sl}=10^{-3}$ a.u. and $E_{sl}=0$ a.u. in benzene and
$E_{sl}=2\times 10^{-3}$ a.u. and $E_{sl}=0$ a.u. in anthracene. 
To the scale of the figure the planar averages calculated by FEF or
by DFPT coincide. 
The error, reported to an enlarged scale in the insets, is always lower 
than $1 \% $ of the value of $\Delta n_{av}$. 

Now we move on to the dielectric properties of the slabs.
We begin by extracting the polarizability of the slab from the induced
dipole moment per unit surface. We define the polarizability of a slab $\alpha$,
starting from the relationship:
\begin{equation}
m = {\alpha \over S} E_{loc},
\label{alpha_slab}
\end{equation}
where $m=-e \int_{-\infty}^\infty \Delta n_{av}(\lambda)\lambda\ d\lambda$ 
is the induced dipole moment per unit surface area 
and $E_{loc}$ is the electric field, perpendicular to the surface,
which induces the dipole moment on the slab ($E_{loc}$ is
the uniform electric field that remains in the vacuum between two slabs 
after subtraction of all short range inhomogeneous fields).
In a periodic slab geometry, the dipole moment of the slab is the origin 
of a sizable electric field. Actually, as described in 
Refs.~\onlinecite{Bengtsson,Vanderbilt2}, in an isolated slab a 
dipole moment per unit surface area produces an the electrostatic 
potential which has different constant values in the vacuum 
on the left and on the right part of the slab. In a repeated slab 
geometry, this jump cannot be accomodated with periodic 
boundary conditions.  
Therefore, an artificial uniform electric field appears in the super-cell 
in order to restore the periodicity of the electrostatic potential. 
Before applying Eq.~\ref{alpha_slab}, this field
has to be added to $E_{sl}$ or to $E$ in order to calculate $E_{loc}$
which actually induces the dipole moment on the slab.
As in Refs.~\onlinecite{Bengtsson,Vanderbilt2} we evaluate $E_{loc}$ as: 
\begin{equation}
E_{loc} = E_{0} + {4 \pi m \over L},  
\label{elocsurf}
\end{equation}
where $L$ is the length of the super-cell in the direction 
perpendicular to the surface and either $E_0=E_{sl}$ (FEF) 
or $E_0=E$ (DFPT).
Using Eqs.~\ref{alpha_slab} and \ref{elocsurf}, we calculate
the polarizability of a slab from the dipole moment per unit surface area as:
\begin{equation}
\alpha= {\Omega \over 4 \pi} {x\over 1+x},
\label{alpha}
\end{equation}
where $x= 4\pi m /(L E_0)$. In Tab.~\ref{tab-pol-slab}, we report
$\alpha$ calculated by Eq.~\ref{alpha} and compare
with $\alpha$ obtained without including the field due to the periodic
boundary conditions ($\alpha=S m/E_0$). It is found that
the polarizability calculated by Eq.~\ref{alpha} is already converged 
at the smallest vacuum space for both benzene and anthracene.
Tab.~\ref{tab-pol-slab} shows that the field due to the
periodic boundary conditions has the same magnitude of the applied 
electric field.
For instance, in benzene (anthracene), at the minimum slab-slab distance, 
for $L=65.0$ a.u. ($L=36.2$ a.u.), $E_{loc}$ is about $123\%$ ($75\%$) larger 
than $E_{0}$. A vacuum size of about $3530$ a.u. ($1530$ a.u.) 
would be necessary 
to reduce the effect of the field due to the boundary conditions below $1 \%$.

Now, we can apply a similar argument and calculate the polarizability 
of a slab using the dielectric constant (Eq.~\ref{eq1}) of the periodic 
solid simulated in the super-cell approach.
If ${\bf n}$ is a unit vector normal to the surface, the
relevant dielectric constant for fields normal to the surface is
$\epsilon=\sum_{\alpha\beta} \epsilon_{\alpha\beta} n_\alpha n_\beta$.
In benzene the surface normal is along the $(010)$ direction and 
$\epsilon=\epsilon_{yy}$, in anthracene it is in the $xy$ plane and 
$\epsilon=\epsilon_{xx}n_x^2+2\epsilon_{xy}n_x n_y+\epsilon_{yy} n_y^2$
where ${\bf n}=(n_x,n_y,0)$.
We report in Tab.~\ref{tab-diel-slab} the dielectric constant 
as a function of the length of the super-cell. 
The dielectric constant depends on the size of the super-cell. 
Instead, the polarizability $\alpha$ becomes constant
as soon as the vacuum size avoids the direct interaction between 
neighbouring slabs.
From Eq.~\ref{eq1}, we have
$\epsilon=1+{4\pi\over \Omega} \alpha E_{loc}/E$ and, by
using Eq.~\ref{elocsurf}, we obtain the expression of the polarizability 
of the isolated slab as a function of the dielectric constant:
\begin{equation}
\alpha = {\Omega\over 4\pi} \left( {\epsilon - 1 \over \epsilon} \right).
\label{final_alpha_slab}
\end{equation}
We report in Tab.~\ref{tab-diel-slab} the polarizability for each vacuum 
distance calculated by Eq.~\ref{final_alpha_slab}.  These values converge 
to the same values of Tab.~\ref{tab-pol-slab} and the convergence rate 
is similar. 

Now, we compare the dielectric properties 
of the surface with those of the bulk. Moreover, we extract the 
dielectric constant of the bulk (in the direction of the surface normal) 
from the slabs polarizabilities.
We use a method inspired by Ref.~\onlinecite{resta}, and restrict our attention 
to the $(010)$ surface of benzene.
For a fixed super-cell size ($L=68.9$ a.u.) we calculate the 
polarizability of benzene slabs with different numbers of atomic layers. 
Fig.~\ref{fig-pol-lay} shows $\alpha(N)$ calculated
as described above for $N=2$ to $N=6$. It is found that $\alpha(N)$
increases linearly with the number of layers. 
We can understand this behavior using Eq.~(5) of Ref.~\onlinecite{resta}.
The polarization $P_N$ of an isolated $N$ layers slab is approximately
\begin{equation}
P_N \approx \sigma_\infty + {2 S p_\infty \over N \Omega_B},
\label{eq-resta}
\end{equation}
where $\sigma_\infty$ is the surface charge of a very thick slab 
in which the surface contribution to the polarization is negligible, 
$ p_\infty$ is the sum of the surface-dipoles which accounts for the
difference between the polarizability of the surface layers and that of
the bulk. $\Omega_B$ is the volume of a bulk unit cell 
with two layers.
In our example, $P_N$ calculated from the polarizability of
the slab is $P_N = 2 \alpha(N) E_{loc}/N\Omega_B$.
$\sigma_\infty$ can be calculated from the bulk dielectric constant. 
The electrostatic of a macroscopic slab in an external field $E_{loc}$ gives:
\begin{equation}
\sigma_\infty = {1\over 4 \pi} {\epsilon_B - 1 \over  \epsilon_B } E_{loc},
\end{equation}
and from Eq.~\ref{eq-resta}, we get:
\begin{equation}
\alpha(N)= N {\Omega_B \over 8 \pi} \left ({\epsilon_B-1 \over \epsilon_B}
\right) + 
{p_\infty S \over E_{loc}}.
\label{final}
\end{equation}
Therefore, the polarizability of an $N$ layers slab is linear in the number
of layers, the slope of the straight line depends only on the 
bulk dielectric constant, and the intercept at the origin 
measures the difference between the dielectric properties of the 
bulk and of the slab surfaces. 
The fit gives $\epsilon_B=2.91$ close to the value of the dielectric 
constant calculated in bulk benzene $\epsilon_{yy}=2.87$. 
The term ${p_\infty S / E_{loc}}$ has the dimensions of a polarizability. 
Defining $\alpha_S={p_\infty S \over 2 E_{loc}}$, where the factor $2$
accounts for two slab surfaces, we obtain $\alpha_S=4.5$ a.u..
Therefore the surface layers are slightly more
polarizable than the bulk, but in benzene the effect is indeed very small.
Of course, this conclusion is valid only for the electronic contribution 
to the dielectric constant, and is obtained in a model system where
the truncated bulk geometry is used for the surfaces.
Finally, we note that, in benzene, the bulk unit cell contains
two layers,  but the polarizability of the slabs does not show any
even-odd effect in Fig.~\ref{fig-pol-lay} for symmetry reasons. 
\begin{table}[t]
\begin{tabular}{cccccc}
 & Benzene & & & Anthracene &   \\
$L$ (a.u.)& $\epsilon$ & $\alpha$ (Eq.~\ref{final_alpha_slab}) & 
$L$ (a.u.) & $\epsilon$ & $\alpha$ (Eq.~\ref{final_alpha_slab}) \\
\hline 
65.0 & 2.2304  &  500.9   & 36.17   & 1.7549   &  298.7  \\
68.9 & 2.0850  &  500.9   & 39.46   & 1.6489   &  298.1  \\
75.0 & 1.9160  &  500.9   & 42.75   & 1.5705   &  298.1  \\
80.0 & 1.8123  &  500.9   & 46.03   & 1.5090   &  298.1  \\
85.0 & 1.7296  &  500.9   & 49.32   & 1.4595   &  298.1  \\
\end{tabular}
\caption{Dielectric constant of the ``solid'' made up of periodically
repeated benzene and anthracene slabs separated by vacuum space, as 
a function of the length ($L$) of the super-cell. The slab polarizability
$\alpha$ (in a.u.) is evaluated using Eq.~\ref{final_alpha_slab}.}
\label{tab-diel-slab}
\end{table}

\section*{Acknowledgment}
We thank E. Tosatti for suggesting the study of benzene and anthracene
surfaces and for useful discussions. 
This project was sponsored by COFIN, by INFM (I.T. Calcolo Parallelo, 
Sezioni F e G, PAIS Chemde) by EU TMR Fulprop, and by MINOS.


\begin{thebibliography}{99}

\bibitem{Vanderbilt}  D. Vanderbilt, Phys. Rev. B {\bf 41}, 7892 (1990).

\bibitem{Laasonen} K. Laasonen, A. Pasquarello, R. Car, C. Lee, and
D. Vanderbilt, Phys. Rev. B {\bf 47}, 10142 (1993).

\bibitem{king-smith} D. Vanderbilt and R.D. King-Smith, Cond-mat/9801177.

\bibitem{smogunov} A. Smogunov, A. Dal Corso, and E. Tosatti, submitted. 
A. Smogunov, A. Dal Corso, and E. Tosatti, Surf. Sci. to appear. 
Cond-mat/0310335.

\bibitem{uno} A. Dal Corso, A. Pasquarello, and A. Baldereschi,
Phys. Rev. B {\bf 56}, R11369 (1997).

\bibitem{due} A. Dal Corso, Phys. Rev. B {\bf 64}, 235118 (2001).

\bibitem{giapponesi} N. Ohba, K. Miwa, N. Nagasako, and A. Fukumoto,
Phys. Rev. B {\bf 63}, 115207 (2001).

\bibitem{baroni} S. Baroni, P. Giannozzi, and A. Testa, Phys. Rev. Lett.
{\bf 58}, 1861 (1987).

\bibitem{rmp} S. Baroni, S. de Gironcoli, A. Dal Corso, and P. Giannozzi,
Rev. Mod. Phys. {\bf 73}, 515 (2001).

\bibitem{baroni2} S. Baroni, P. Giannozzi, and A. Testa, Phys. Rev. Lett.
{\bf 59}, 2662 (1987).

\bibitem{Sprik2} E.J. Meijer and M. Sprik, J. Chem. Phys. {\bf 105}, 8684
(1996).

\bibitem{Hummer} K. Hummer, P. Puschnig, and C. Ambrosch-Draxl,
Phys. Rev. B {\bf 67}, 184105 (2003).

\bibitem{wyckoff} Wyckoff, {\em Crystal structures v.6: The structure 
of benzene derivatives. 2nd ed.}, Wiley, 1969-1971.

\bibitem{KR1} K. Kunc and R. Resta, Phys. Rev. Lett. {\bf 51}, 686 (1983).

\bibitem{KR2} R. Resta and K. Kunc, Phys. Rev. B {\bf 34}, 7146 (1986).

\bibitem{Nunes} R.W. Nunes and D. Vanderbilt, Phys. Rev. Lett. {\bf 73},
712 (1994).

\bibitem{Fernandez} P. Fern\`andez, A. Dal Corso, and A. Baldereschi,
Phys. Rev. B {\bf 58}, R7480 (1998).

\bibitem{Vanderbiltef} I. Souza, J. \'I\~niguez, and D. Vanderbilt,
Phys. Rev. Lett. {\bf 89}, 117602 (2002).

\bibitem{Pasquarello} P. Umari and A. Pasquarello, Phys. Rev. Lett. 
{\bf 89}, 157602 (2002).

\bibitem{Umari} P. Umari, X. Gonze, and A. Pasquarello, Phys. Rev. Lett.
{\bf 90}, 027401 (2003).

\bibitem{PWSCF}S. Baroni, A. Dal Corso, S. de Gironcoli, and P. Giannozzi, 
{\tt http://www.pwscf.org}.

\bibitem{PBE} J.P. Perdew, K. Burke, and M. Ernzerhof, Phys. Rev. Lett. 
{\bf 77}, 3865 (1996).

\bibitem{MP}
H. J. Monkhorst and J. D. Pack, Phys. Rev. B {\bf 13}, 5188 (1976).

\bibitem{JCP-Dele} M.S. Deleuze, A.B. Trofimov, and L.S. Cederbaum, 
J. Chem. Phys. {\bf 115}, 5859 (2001).

\bibitem{Ketkar} S.N.~Ketkar, M.~Kelley, M.~Fink, and R.Ch.~Ivey, J.~Mol.~Struct.~{\bf 77}, 127 (1981).

\bibitem{an-cm} For a more accurate formula see also
K. Urano and M. Inoue, J. Chem. Phys. {\bf 66}, 791 (1977). 

\bibitem{pol-mp2}E. Perrin, P.N. Prasad, P. Mougenot, and M. Dupuis, 
J. Chem. Phys. {\bf 91}, 4728 (1989).

\bibitem{pol-dft} A.A. Quong and M.R. Pederson, Phys. Rev. B 
{\bf 46}, 12906 (1992).

\bibitem{pol-nmr} B.H.~Ruessink and C.~MacLean, 
J. Chem. Phys. {\bf 85}, 93 (1986).
The experimental value $231$ a.u. ($174$, $367$, and $154$ for 
$\alpha_{xx}$,$\alpha_{yy}$ and $\alpha_{zz}$ respectively) 
for the mean polarizability of anthracene reported in Tab.~II of 
this reference seems too different
from the other values and has not been reported in the text and in 
the Table.

\bibitem{noi} J. Tobik, A. Dal Corso, S. Scandolo and E. Tosatti, Surf. Sci.
To appear.

\bibitem{Bengtsson} L. Bengtsson, Phys. Rev. B {\bf 59}, 12301 (1999).

\bibitem{Vanderbilt2} B. Meyer and D. Vanderbilt, Phys. Rev. B {\bf 63},
205426 (2001). 

\bibitem{resta} L. Fu, E. Yaschenko, L. Resca, and R. Resta, 
Phys. Rev. B {\bf 60}, 2697 (1999).

\bibitem{Soos} Z.G. Soos, E.V. Tsiper, and R.A. Pascal Jr., 
Chem. Phys. Lett. {\bf 342}, 652 (2001).

\bibitem{Reis} H. Reis, M.G. Papadopoulos, P. Calaminici, K. Jug, 
and A.M. K\"{o}ster, Chem. Phys. {\bf 261}, 359 (2000).

\bibitem{Cheng-AJC} C.L. Cheng, D.S.N. Murthy, and G.L.D. Ritchie, 
Aust. J. Chem. {\bf 25}, 1301 (1972).

\bibitem{LeFevre} R.J.W. Le F\`{e}vre and G.L.D. Ritchie, 
J. Chem. Soc. {\bf B 1968}, 775.

\bibitem{Vuks} M.F. Vuks, Opt. Spectrosc. {\bf 20}, 361 (1966).

\bibitem{Bounds-ChP} P.J. Bounds and R.W. Munn, 
Chem. Phys. {\bf 24}, 343 (1977).

\bibitem{Perez-JMS} J.J. Perez, A.J. Sadlej, J. Mol. Struct. 
(Theochem) {\bf 371}, 31 (1996).

\end{thebibliography}
\end{document}